\documentclass[twocolumn]{aastex63}

\graphicspath{{./}{figures/}}
\usepackage{amsmath}
\begin{document}

\title{Neutron Star Mergers in AGN Accretion Disks: Cocoon and Ejecta Shock Breakouts}

\author[0000-0002-9195-4904]{Jin-Ping Zhu}
\affil{Department of Astronomy, School of Physics, Peking University, Beijing 100871, China; \url{zhujp@pku.edu.cn}}

\author[0000-0002-9725-2524]{Bing Zhang}
\affiliation{Department of Physics and Astronomy, University of Nevada, Las Vegas, NV 89154, USA; \url{zhang@physics.unlv.edu}}

\author[0000-0002-1067-1911]{Yun-Wei Yu}
\affiliation{Institute of Astrophysics, Central China Normal University, Wuhan 430079, China; \url{yuyw@mail.ccnu.edu.cn}}

\author[0000-0002-3100-6558]{He Gao}
\affiliation{Department of Astronomy, Beijing Normal University, Beijing 100875, China; \url{gaohe@bnu.edu.cn}}

\begin{abstract}
Neutron star mergers are believed to occur in accretion disks around supermassive black holes. Here we show that a putative jet launched from the merger of a binary neutron star (BNS) or a neutron star--black hole (NSBH) merger occurring at the migration trap in an active galactic nucleus (AGN) disk would be choked. The jet energy is deposited within the disk materials to power a hot cocoon. The cocoon is energetic enough to break out from the AGN disk and produce a bright X-ray shock breakout transient peaking at $\sim0.15\,{\rm d}$ after the merger. The peak luminosity is estimated as $\sim 10^{46}\,{\rm erg}\,{\rm s}^{-1}$, which can be discovered by Einstein Probe from $z\lesssim 0.5$. Later on, the non-relativistic ejecta launched from the merger would break out the disk, powering an X-ray/UV flare peaking at $\sim 0.5\,{\rm d}$ after the merger. This second shock breakout signal may be detected by UV transient searches. The cocoon cooling emission and kilonova emission are outshone by the disk emission and difficult to be detected. Future joint observations of gravitational waves from BNS/NSBH mergers and associated two shock breakout signatures can provide a strong support for the compact binary coalescence formation channel in AGN disks.

\end{abstract}


\section{Introduction} \label{sec:intro}

Since the Laser Interferometer Gravitational wave (GW) Observatory (LIGO) detected the first GW signal GW150914 from a binary black hole (BBH) merger \citep{abbott2016observation}, dozens of BBH GW events have been reported \citep{abbott2019gwtc1,abbott2020gwtc2}. The formation channels for BBH systems are still subject to debated. Two main channels include field isolated binary evolution \citep[e.g.,][]{belczynski2010,demink2016,santoliquido2020} and dynamical interactions in dense environments including globular clusters \citep[e.g.,][]{sigurdsson1993,rodriguez2015}, galactic nuclei \citep[e.g.,][]{antonini2016,fragione2019}, and active galactic nucleus (AGN) disks \citep[e.g.,][]{mckernan2012,mckernan2014,bartos2017,stone2017}. BBH mergers can hardly generate electromagnetic (EM) counterparts in the absence of gas\footnote{\cite{zhang2016} suggested that if at least one BH in the BBH systems carries a certain amount of charge, the inspiral of a BBH merger could form a global magnetic dipole. The rapid increase of the global magnetic dipole radiation power could drive a short-duration EM counterpart (e.g. a fast radio burst or even a short-duration GRB) at the coalescence.}. In a dense environment, especially in an AGN accretion disk, BBH mergers are expected to produce EM emission. \cite{mckernan2019} suggested that the GW kick at the remnant BH after a BBH merger can cause ram-pressure stripping of gas within the BH Hill sphere to power a UV/optical EM flare in an AGN disk. Recently, the LIGO/Virgo Collaboration announced a high-mass BBH merger system \citep{abbott2020GW190521}, GW190521, with two BH masses $85^{+21}_{-14}\,M_\odot$ and $66^{+17}_{-18}\,M_\odot$, respectively. Intriguingly, \cite{graham2020} reported a plausible optical EM counterpart ZTF19abanrhr detected by the Zwicky Transient Facility \citep[ZTF;][]{graham2020} $\sim34\,{\rm days}$ after the GW190521 trigger, which was consistent with a plausible association of the GW event with the AGN J$124942.3+344929$. Conversely, \cite{ashton2020} and \cite{nitz2020} argued that the localisation overlap between ZTF19abanrhr with GW190521 is insufficient to confidently associate the two events.

Besides BBH mergers, a large population of binary neutron star (BNS) and neutron star--black hole (NSBH) mergers are also expected to occur in AGN disks \citep{cheng1999,mckernan2020}. These neutron star mergers have long been proposed as the progenitors of short-duration gamma-ray bursts \citep[sGRBs;][]{paczynski1986,paczynski1991,eichler1989,narayan1992} and kilonovae powered by the radioactive decay of {\em r}-process nuclei from the sub-relativistic ejecta of these events \citep{li1998,metzger2010}. The multi-messenger observations of a BNS merger GW event \citep[GW170817;][]{abbott2017GW170817} and its associated EM signals, including a sGRB \citep[GRB\,170817A;][]{abbott2017gravitational,goldstein2017,zhang2018}, a broad-band off-axis jet afterglow \citep[e.g.,][]{margutti2017,troja2017,lazzati2018,lyman2018,ghirlanda2019} and a kilonova \citep[AT\,2017gfo;][]{abbott2017multimenssenger,arcavi2017,coulter2017,drout2017,evans2017,kasliwal2017,pian2017,smartt2017}, provided a smoking-gun evidence for the long-hypothesized origin of sGRBs and kilonovae. The properties of sGRBs and kilonovae generated from BNS and NSBH mergers have been studied in detail. However, no study has focused on the properties of EM counterparts of BNS or NSBH merger systems that are in AGN disks\footnote{\cite{perna2020} recently studied the general properties of GRBs in an AGN disk.}. A smoking-gun signature of the EM counterparts from BNS or NSBH mergers in an AGN disk, if identified, would lend support to the formation channel of compact binary mergers in AGNs. The motivation of this paper is to study the observational signatures of neutron star mergers in an AGN disk.

\section{Cocoon and Ejecta Shock Breakouts} \label{sec:shock}

\begin{figure*}
    \centering
    \includegraphics[width = 1\textwidth , trim = 0 0 0 0, clip]{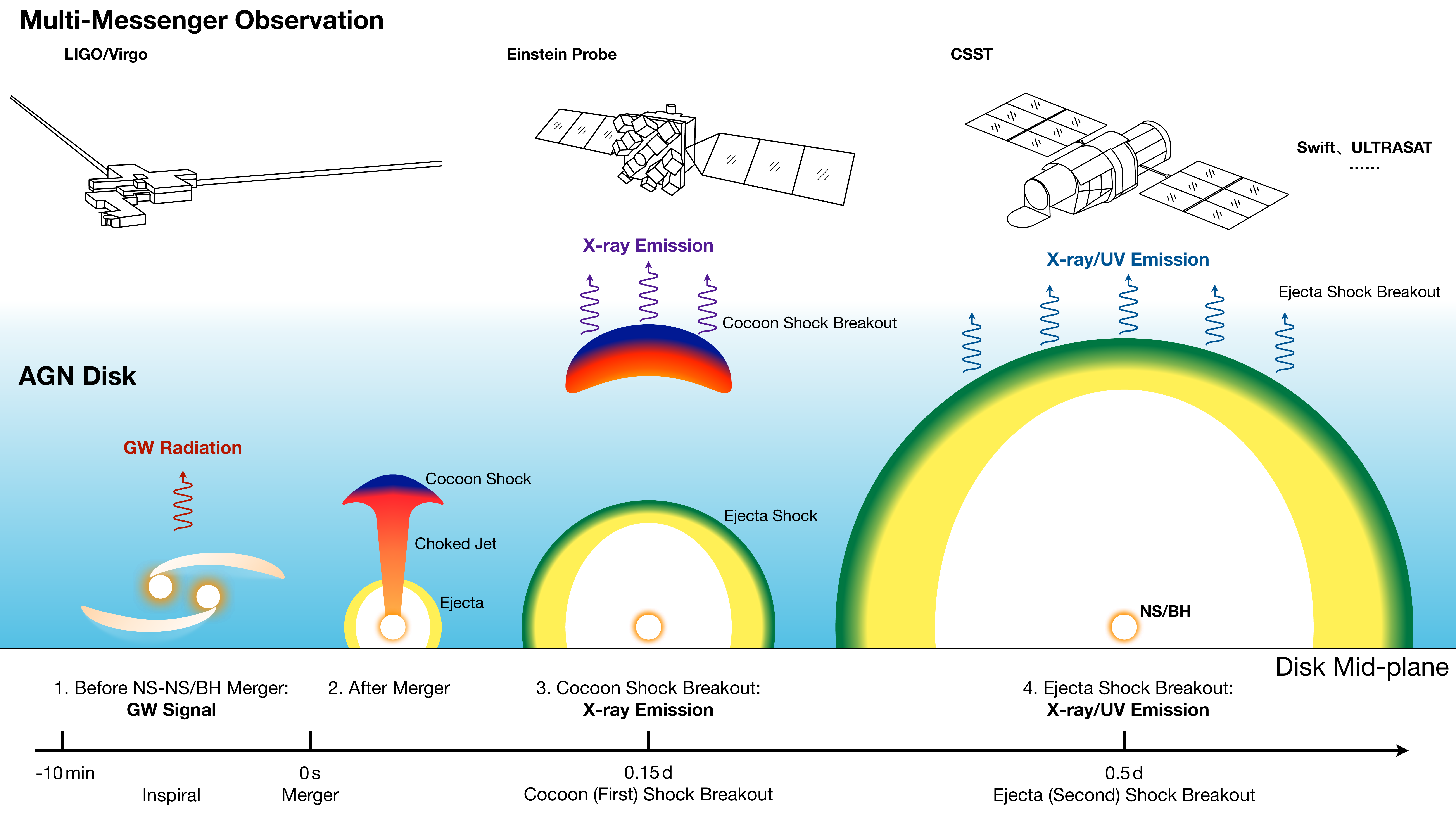}
    \caption{Cartoon picture of a BNS/NSBH merger in an AGN disk.} 
    \label{fig:cartoon}
\end{figure*}

Figure \ref{fig:cartoon} illustrates the physical processes after a neutron star merger (BNS or NSBH merger) in an AGN disk. We consider a neutron star merger in an an AGN disk, assume that merger can launch a relativistic jet, and study its propagation in the disk. For simplicity, we assume that the binary orbital plane is parallel to the disk plane so that the post-merger jet launched from the system is perpendicular to the disk.

A successful jet can penetrate through the surrounding environment (e.g. the stellar envelope of a massive star or the ejecta launched from a neutron star merger) and is surrounded by a hot cocoon with a lower speed \citep[e.g.][]{meszaros2001,ramirez-ruiz2002,zhang2004,nakauchi2013,nakar2017,hamidani21}. In the case of a choked jet, the jet energy from a neutron star merger would be stored in a hot, shocked material, forming a broad cocoon without a spine jet \citep[e.g.,][]{gottlieb2018,piran2019}. As we show below, the GRB jet in an AGN disk cannot penetrate the disk so produce a transrelativistic cocoon of this type. The breakout of this cocoon from the AGN disk would produce X-rays. The slower ejecta launched from the merger can also drive a non-relativistic shock, which breaks out the disk at a later time and emits X-ray/UV photons. 

Compact binary mergers are expected to occur in the migration traps \citep{bellovary2016} plausibly located at $a \sim 10^3\,r_{\rm g}$, where $a$ is the orbital semi-major axis of the system in units of super-massive BH's gravitational radius $r_{\rm g} \equiv GM_{\rm SMBH} / c^2$. The disk height at the location is
\begin{equation}
    H = 1.5\times10^{14}\, M_{\rm SMBH,8\odot}\left(\frac{H/a}{0.01}\right)\left(\frac{a}{10^3\,r_{\rm g}}\right)\,{\rm cm},
\end{equation}
where $M_{\rm SMBH}$ is distributed in $\sim[10^7 , 10^9]\,M_\odot$ based on AGN observations \citep[e.g.,][]{woo2002,kollmeier2006}, and $H/a\sim[10^{-3} , 0.1]$ is the disk aspect ratio \citep{sirko2003,thompson2005}. Therefore, the disk height at migration traps can be $O(10^{14})\,{\rm cm}$. Hereafter the convention $Q_x = Q/10^x$ is adopted in cgs units.

With the consideration of a gas-pressure-dominated disk, its vertical atmospheric density \citep{netzer2013} is $\rho(z) = \rho_0\exp(-z/H)$, where $\rho_0$ is the mid-plane density at migration traps and $z$ is the vertical distance. Near the migration traps, the mid-plane density is almost $\rho_0\sim O(10^{-10})\,{\rm g}\,{\rm cm}^{-3}$ \citep{sirko2003,thompson2005}. The propagation direction and the shape of the shocks can be affected if there is an asymmetric density distribution for the AGN disk atmosphere. For an atmosphere with an exponentially decaying density profile, the density is $\rho\approx\rho_0$ for $z \leq H$ and decreases rapidly for $z \geq H$. For simplicity, hereafter, we assume that the AGN disk has a uniform density profile at $z \leq H$.

\subsection{Cocoon Shock Breakout} \label{sec:cocoon}

After a neutron star merger (BNS or NSBH merger), a pair of relativistic jets would be launched along the axis of the binary merger, which could be observed as an sGRB if the merger is not in the AGN disk and if one of the jet points toward Earth. Observationally, sGRBs have durations shorter than $\sim2\,{\rm s}$ with a distribution peaking at  $t_{\rm j}\approx T_{90}\sim0.8\,{\rm s}$ \citep{kouveliotou1993,horvath2002,nakar2007,berger2014}. The statistic analysis of 103 sGRB broadband afterglow observations by \cite{fong2015} revealed that the median beaming-corrected energy for sGRB jets is $E_{\rm j} \sim 1.6 \times 10^{50}\,{\rm erg}$. Therefore, the jet luminosity $L_{\rm j} \approx E_{\rm j}/t_{\rm j}$ is usually a few $10^{50}\,{\rm erg}\,{\rm s}^{-1}$. 

\begin{figure}
    \centering
    \includegraphics[width = 1\linewidth , trim = 50 30 40 30, clip]{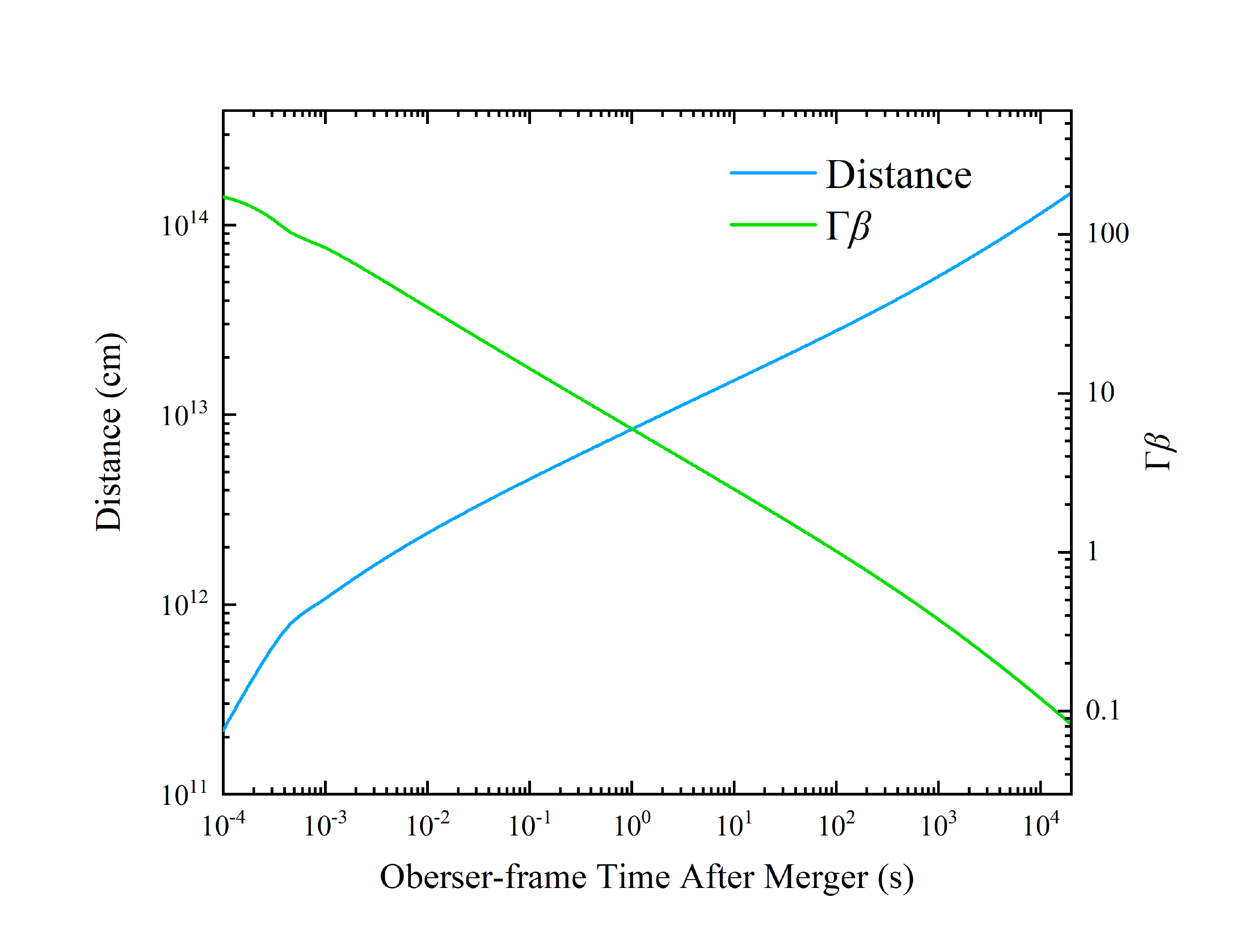}
    \caption{Evolution of distance and $\Gamma\beta$ for a jet moving in an AGN disk. The following parameters are adopted: $E_{\rm j}= 10^{50}\,{\rm erg}$, $\theta_{\rm j} = 10^\circ$,  $\Gamma_{\rm j} = 200$, and $n \approx \rho_{0,-10}/m_p = 6\times10^{13}\,{\rm cm}^{-3}$.}
    \label{fig:JetEvolution}
\end{figure}

The jet initially penetrates through the merger ejecta and get collimated there. The jet breaks out from the ejecta in a $\sim0.1\,{\rm s}$, which shorter than the duration of sGRB \citep[e.g.,][]{yu2020}. When it enters the AGN disk with a much lower density, it loses collimation but is re-accelerated by the waste heat in the ejecta cocoon. Energy from the central engine would be continually injected into the reborn fireball and reach a relativistic speed. Following \cite{bromberg2011a}, the critical parameter that determines the evolution of the jet is $\tilde{L} \simeq L_{\rm j} / \rho_0c^3(\pi r_{\rm j}^2) \approx (10 ^ {14} \,{\rm cm} / r_{\rm j}) ^ 2$. Since the jet radius $r_{\rm j} \ll 10^{\rm 14}\,{\rm cm}$, one has $\tilde{L} \gg 1$ so that the jet head would travel with a relativistic speed. The jet would be uncollimated and then sweep AGN material to decelerate significantly. One may calculate the dynamical evolution of the jet to check whether the jet can successfully break out from the disk. For an order of magnitude estimation, hereafter we assume that the merger is located at the disk mid-plane and the propagation direction of the jet is perpendicular to the disk surface. The dynamical equation covering both ultra-relativistic and non-relativistic shock dynamics may be written as \citep{huang1999}
\begin{equation}
    \frac{d\Gamma}{dt} = -\frac{\Gamma ^ 2 - 1}{M_{\rm j} + 2\Gamma m_{\rm sw}},
\end{equation}
where $\Gamma$ is the bulk Lorentz factor of the external shock, $M_{\rm j}$ is the jet mass, and $m_{\rm sw}$ is the swept-up disk medium mass. The total energy of the jet is $E_{\rm j} = \Gamma_{\rm j}M_{\rm j}c^2$ with $\Gamma_{\rm j}$ is the jet Lorentz factor after breaking out the merger ejecta. The evolution of the swept-up disk mass $m_{\rm sw}$ and the distance $r$ traveled by the jet can be calculated by
\begin{equation}
\begin{split}
    \frac{dm_{\rm sw}}{dr} &= 2\pi r^2(1 - \cos\theta_{\rm j})nm_p, \\
    \frac{dr}{dt} &= \frac{\beta c}{1 - \beta},
\end{split}
\end{equation}
where $\theta_{\rm j}$ is the half opening angle of the jet, $n \approx \rho_0/m_p$ is the disk number density, $m_p$ is the proton mass, and $\beta = \sqrt{1 - \Gamma^{-2}}$. As shown in Figure \ref{fig:JetEvolution}, the jet in an AGN disk is quickly decelerated significantly before escaping the disk. The jet can be choked if the jet tail catches up the jet head. This requires $(v_{\rm t} - v_{\rm h})t_{\rm j,bo} > z_{\rm j}$, where $v_{\rm t} \approx c$ is the velocity of the jet tail, $v_{\rm h}$ is the velocity of the jet head, $t_{\rm j,bo}$ is the time for the jet to break out from the AGN disk, and $z_{\rm j} \approx{ct_{\rm j}}$ is the jet length. So the condition for the jet to be choked is $t_{\rm j,bo}>2\Gamma_{\rm j}^2t_{\rm j}$. Since the jet is decelerated to a trans-relativistic speed in the AGN disk as shown in Figure \ref{fig:JetEvolution}, i.e., $\Gamma_{\rm j} \approx 1-2$, one would always have $t_{\rm j,bo} \gg t_{\rm j}\approx 1\,{\rm s}$. Therefore, the jet would be stalled and deposit its energy to the cocoon with the cocoon energy is $E_{\rm c}\approx E_{\rm j}$. The resulting profile of the cocoon is shaped like a cone with a half-opening angle $\theta_{\rm c}$ that is usually larger than $\theta_0\sim10^\circ$ \citep{bromberg2011a,bromberg2011b}. The hot, transrelativistic cocoon expands, forms a radiation-mediated shock, sweeps the AGN disk material, and finally breaks out from the disk. 

The shock would break out if the photons ahead of the shock diffuse faster than the shock propagation. The photons from the vertical location $h$ would take $t_{\rm diff} \approx \kappa\rho_0(H-h)^2/c$ to diffuse to the disk surface at $H$ , where $\kappa$ is the opacity. The shock propagation time is $t_{\rm exp}\approx(H - h) / v_{\rm c}$, where $v_{\rm c}$ is the velocity of the faster part of the cocoon shock. The shock breakout takes place when $t_{\rm diff}\approx t_{\rm exp}$, so we can calculate the distance between the location of shock breakout and disk surface as
\begin{equation}
\label{eq:BreakoutLocation}
    d = H-h \approx 3\times10^{11}\,\rho_{0,-10}^{-1}\left(\frac{v_{\rm c}}{0.1\,c}\right)^{-1}\,{\rm cm},
\end{equation}
where $\kappa=0.34\,{\rm cm^2}\,{\rm g}^{-1}$ is adopted. Since $d\ll H$, the shock breakout would take place near the AGN disk surface. 

When the shock breaks out, the total mass of the swept material by the cocoon is 
\begin{equation}
    M_{\rm c,sw} \approx \frac{\pi\rho_0\theta_{\rm c}^2H^3}{3} = 1.6\times 10^{-3}\,\rho_{0,-10}\theta_{\rm c,10^\circ}^2H_{14}^3\,M_\odot.
\end{equation}
The shock breakout velocity can be estimated as $v_{\rm c}\approx\sqrt{E_{\rm c}/M_{\rm c,sw}}$ which is of the order
\begin{equation}
    v_{\rm c}/c \approx 0.19\,E_{\rm c,50}^{1/2}\rho_{0,-10}^{-1/2}\theta_{\rm c,10^\circ}^{-1}H_{14}^{-3/2}.
\end{equation}

Because $v_{\rm c}/c\gtrsim0.1$, there is not enough time to form a photon-electron thermal equilibrium and a blackbody spectrum \citep{weaver1976,katz2010K,nakar2010}. In such a case, electrons and photons are in Compton equilibrium at a much higher temperature $T_{\rm c}$ than the downstream temperature $T_{\rm c}^{\rm BB}$. With an adiabatic index of $\gamma = 4 / 3$ and by comparing the radiation energy density with the kinetic energy density, one can express the ejecta shock downstream temperature as $T_{\rm c}^{\rm BB} \approx (7\rho_0v_{\rm c}^2/2a)^{1/4}$, where $a$ is the radiation constant. Therefore, the downstream temperature is $T_{\rm c}^{\rm BB} \approx 1.1\times10^6\,E_{\rm c,50}^{1/4}\theta_{\rm c,10^\circ}^{-1/2}H_{14}^{-3/4}\,{\rm K}$. \cite{katz2010K} suggested that $T_{\rm c}$ can be modified by Comptonization as $T_{\rm c}=T_{\rm c}^{\rm BB} \eta^2/\xi(T_{\rm c})^2$, where $\eta \approx 16\,\rho_{0,-10}^{-1/8}(v_{\rm c}/0.1\,c)^{15/4}$ is the thermal coupling coefficient of the breakout shell at the initial time in the expanding gas, and $\xi(T_{\rm c}) \approx {\rm max} [1 , 0.5\ln(y_{\rm max})(1.6 + \ln(y_{\rm max})]$ is the Comptonization correction factor. Here, $y_{\rm max} = k_{\rm B}T_{\rm c}/h\nu_{\rm min}\approx1.2\times10^{3}\,\rho_{0,-10}^{-1/2}T_{\rm c,7}^{9/4}$ is the ratio between $T_{\rm c}$ and the lowest photon energy for Comptonization, where $k_{\rm B}$ is the Boltzmann constant. For $v_{\rm c}/c = 0.19$ and $\rho_0 = 10^{-10}\,{\rm g}\,{\rm cm}^{-3}$, one can solve $T_{\rm c}\approx1.8\times10^{7}\,{\rm K}$ while the observed temperature is $T_{\rm c}^{\rm obs} \approx (v_{\rm c}/c)^{1/4}T_{\rm c} \approx 1.2\times10^{7}\,{\rm K}$. Therefore, the main radiation of the cocoon shock breakout is in the soft X-ray band.

With the breakout velocity known, the distance from the AGN disk surface to the shell in which the breakout occurs can be estimated by Equation (\ref{eq:BreakoutLocation}), i.e., $d_{\rm c,bo} \approx 1.6\times10^{11}\,E_{\rm c,50}^{-1/2}\rho_{0,-10}^{1/2}\theta_{\rm c,10^\circ}H_{14}^{3/2}\,{\rm cm}$, while the diffusion time is 
\begin{equation}
    t_{\rm c,diff} \approx \frac{\kappa\rho_0d_{\rm c,bo}^2}{c} \approx28\,E_{\rm c,50}^{-1}\theta_{\rm c,10^\circ}^2H_{14}^3\,{\rm s}.
\end{equation}
The energy released during the shock breakout is approximately $E_{\rm c,bo} \approx m_{\rm c}v_{\rm c}^2$, where $m_{\rm c} \approx \rho_0\pi\theta_{\rm c}^2H^2d_{\rm c,bo}$ is the mass of the breakout layer, i.e.,
\begin{equation}
    E_{\rm c,bo} \approx 4.7\times10^{47}\,E_{\rm c,50}^{1/2}\rho_{0,-10}^{-1/2}\theta_{\rm c,10^\circ}H_{14}^{1/2}\,{\rm erg}.
\end{equation}
Therefore, the luminosity of the cocoon shock breakout is approximately
\begin{equation}
    L_{\rm c,bo}\approx\frac{E_{\rm c,bo}}{t_{\rm c,diff}} \approx 1.7\times10^{46}\,E_{\rm c,50}^{3/2}\rho_{0,-10}^{-1/2}\theta_{\rm c,10^\circ}^{-1}H_{14}^{-5/2}\,{\rm erg}\,{\rm s}^{-1}.
\end{equation}
One can also estimate the cocoon breakout time after the merger as $t_{\rm c,bo} \approx H / \sqrt{2}v_{\rm c}$, i.e.,
\begin{equation}
    t_{\rm c,bo} \approx 0.15\,E_{\rm c,50}^{-1/2}\rho_{0,-10}^{1/2}\theta_{\rm c,10^\circ}H_{14}^{5/2}\,{\rm d}.
\end{equation}

\subsection{Ejecta Shock Breakout}

The comprehensive observations \citep{cowperthwaite2017,kasen2017,kasliwal2017,perego2017,tanaka2017,villar2017} for GW170817/AT\,2017gfo indicated that the total mass of the ejecta lies in the range of $\sim0.04-0.08M_\odot$ with a median velocity $\sim0.1-0.2\,c$. Therefore, the total kinetic energy of the ejecta $E_{\rm e}$ could only be a few $10^{51}\,{\rm erg}$. Similar to the cocoon, the non-relativistic ejecta can also sweep AGN disk materials, form an ejecta shock, and finally breakout from the disk.

We assume that the aforementioned cocoon shock would not change the disk environment significantly, and the ejecta profile formed after BNS and NSBH mergers is symmetric. The ejecta shock would also break out near the disk surface. The total mass of the swept material by the ejecta shock is
\begin{equation}
    M_{\rm sw,e}\approx\frac{4\pi\rho_0 H^3}{3} = 0.21\,\rho_{0, -10}H_{14}^3\,M_\odot.
\end{equation}
The ejecta shock velocity upon breakout can be estimated as $v_{\rm e}\approx\sqrt{E_{\rm e}/(M_{\rm e,sw} + M_{\rm ej})}$, where $M_{\rm ej}$ is the ejecta mass. Since $M_{\rm e,sw}\gg M_{\rm ej}$, $v_{\rm e}$ is approximately
\begin{equation}
    v_{\rm e}/c \approx 0.05\,E_{\rm e,51}^{1/2}\rho_{0,-10}^{-1/2}H_{14}^{-3/2}.
\end{equation}
Because $v_{\rm e}/c\lesssim0.1$, different from the cocoon shock, there is no significant departure from thermal equilibrium at ejecta shock breakout so that its emission spectrum can be roughly represented by a blackbody spectrum. The temperature is given by
\begin{equation}
    T_{\rm e} \approx 6.9\times10^5\,E_{\rm e,51}^{1/4}H_{14}^{-3/4}\,{\rm K},
\end{equation}
which is mainly dependent on the ejecta kinetic energy and the disk height. The observed temperature is
\begin{equation}
T_{\rm e}^{\rm obs} \approx \left(\frac{v_{\rm e}}{c}\right)^{1/4}T_{\rm e} \approx 2.7\times10^5\,E_{\rm e,51}^{3/8}\rho_{0,-10}^{-1/8}H_{14}^{-9/8}\,{\rm K}.
\end{equation}
Therefore, the ejecta shock breakout signal would be a UV/X-ray flash.

Similar to the cocoon shock breakout, the energy released is $E_{\rm e,bo}\approx m_{\rm e}v_{\rm e}^2$, where $m_{\rm e} \approx \rho_0\pi H^2d_{\rm bo,e}$, i.e.,
\begin{equation}
E_{\rm e,bo} \approx 4.3\times10^{48}\,E_{\rm e,51}^{1/2}\rho_{0,-10}^{-1/2}H_{14}^{1/2}\,{\rm erg}.
\end{equation}
The diffusion time is
\begin{equation}
    t_{\rm e,diff} \approx370\,E_{\rm e,51}^{-1}H_{14}^3\,{\rm s}.
\end{equation}
The luminosity of the ejecta shock breakout can be then estimated as
\begin{equation}
    L_{\rm e,bo}\approx\frac{E_{\rm e,bo}}{t_{\rm e,diff}} \approx 1.2\times10^{46}\,E_{\rm e,51}^{3/2}\rho_{0,-10}^{-1/2}H_{14}^{-5/2}\,{\rm erg}\,{\rm s}^{-1}.
\end{equation}
The breakout time after the merger is $t_{\rm e,bo} \approx H / \sqrt{2}v_{\rm e}$, i.e.,
\begin{equation}
    t_{\rm e,bo} \approx 0.53\,E_{\rm e,51}^{-1/2}\rho_{0,-10}^{1/2}H_{14}^{5/2}\,{\rm d}.
\end{equation}

In summary, if there is a neutron star merger in an AGN disk, two shock breakout events, i.e. a cocoon breakout and an ejecta breakout, would occur. Both shock breakouts have similar luminosities of the order of $10^{46}\,{\rm erg}\,{\rm s}^{-1}$. The cocoon shock breaks out earlier at $\sim0.15\,{\rm d}$ after the merger, while the ejecta shock breakout time is $\sim0.5\,{\rm d}$. We can also conclude the cocoon shock breakout should be an X-ray transient, while the ejecta shock breakout, having a low observed temperature, would be a UV/X-ray transient.

\section{Implication for Future Searches} \label{sec:detection}

\subsection{Comparison with the AGN disk spectra}

\begin{figure}
    \centering
    \includegraphics[width = 1\linewidth , trim = 50 30 80 10, clip]{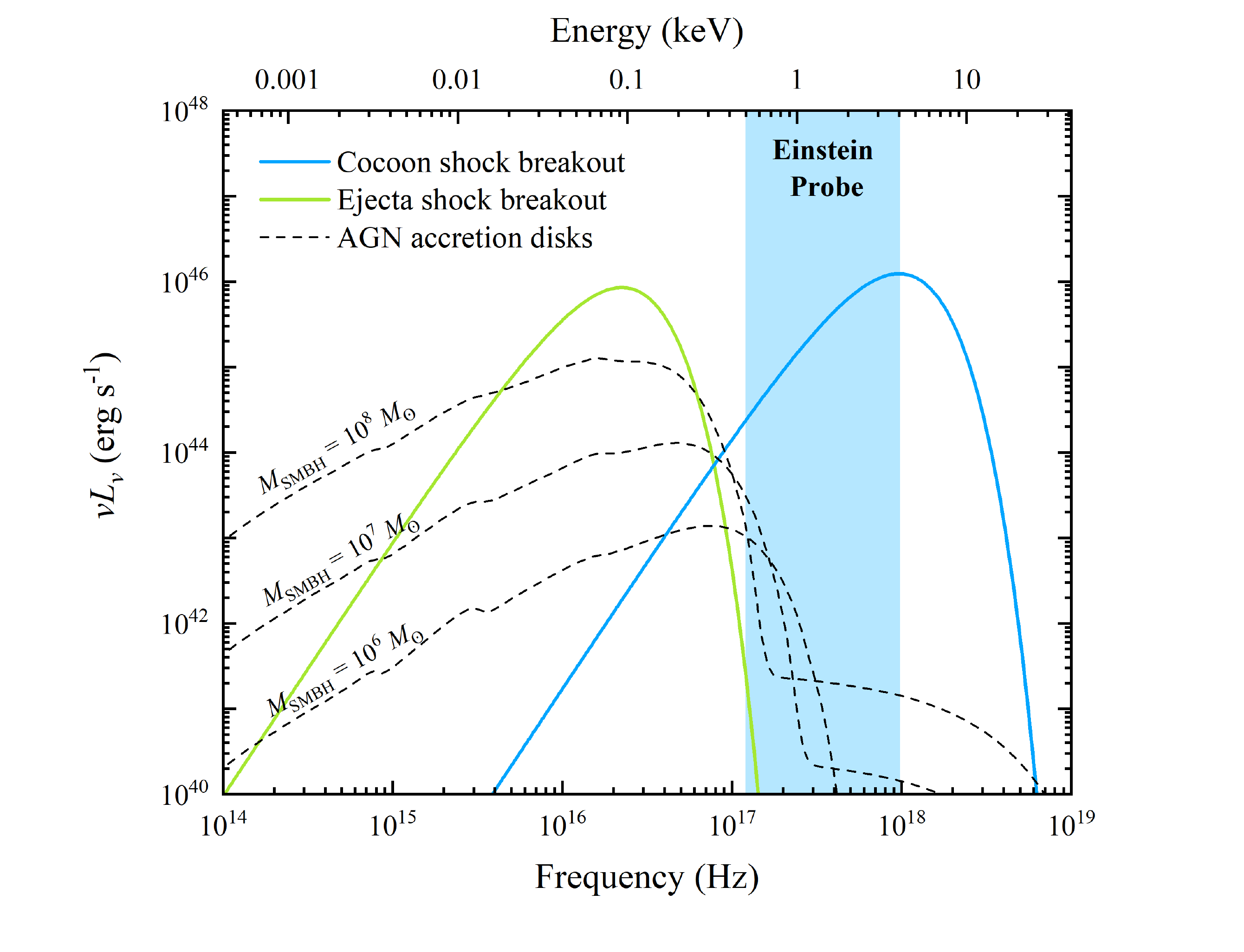}
    \caption{Spectra at the cocoon (blue solid line) and ejecta (green solid line) shock breakout. The dashed lines represent calculated spectrum from SMBHs accreting at nearly the Eddington luminosity ($L/L_{\rm Edd}\sim 0.3$) and central SMBH masses of $M_{\rm SMBH} = 10^6$, $10^{7}$, and $10^8\,M_\odot$ \citep{hubeny2001}. The disk viewing angle is set to $60^\circ$ and viscosity parameter $\alpha = 0.01$. The blue shaded region represent the nominal detection bandpass for the Wide-field X-ray Telescope of the Einstein Probe mission \citep{yuan2016}. } 
    \label{fig:AGNSpectrum}
\end{figure}

We show the indicative spectra of cocoon shock breakout, ejecta shock breakout, AGN accretion disk and its corona \citep{hubeny2001} in Figure \ref{fig:AGNSpectrum}. Various parameters are taken as their typical values (see Figure \ref{fig:AGNSpectrum} caption). Since the X-ray emission from AGNs is less prominent, the cocoon shock breakout signal is less contaminated by the contribution of AGN emission and, therefore, easier to detect. The emission energy of the first (cocoon) breakout lies in the range of the nominal detection bandpass of the Einstein Probe (EP) mission \citep{yuan2016}. After the peak time $\sim 0.15\,{\rm d}$, the temperature and luminosity of the cocoon shock breakout emission would drop rapidly. After $\sim 0.3\,{\rm d}$, the second (ejecta) shock breakout emission, whose temperature is much lower, will emerge. As shown in Figure \ref{fig:AGNSpectrum}, the breakout of the second shock is likely greatly contaminated by the AGN disk emission and outshone by the bright emission from AGN disks in the optical ($3.4-7.9\times10^{14}\,{\rm Hz}$) band. Nonetheless, this component would show up as a bright flare on top of the AGN disk emission in the NUV band, and could be detected by UV ($\sim1-2\times10^{15}\,{\rm Hz}$) transient searches, e.g. by Swift \citep{gehrels2004}, Chinese Space Station Telescope \citep[CSST;][]{zhan2011}, Ultraviolet Transient Astronomy Satellite \citep[ULTRASAT;][]{sagiv2014}, and others. 

Cooling of the shock heated AGN material can be described based on the Appendix C of \cite{piro2018}. The diffusion timescale of cocoon cooling is $t_{\rm cc,diff}\approx(\kappa\rho_0 H^3/3v_{\rm c}c)^{1/2}\approx3\,E_{\rm c,50}^{-1/4}\rho_{0,-10}^{3/4}\theta_{\rm c,10^\circ}^{1/2}H_{14}^{9/4}\,{\rm d}$, while the resulting peak luminosity is 
\begin{equation}
\begin{split}
    L_{\rm cc,p}&\approx\left(\frac{\theta_{\rm c}^2}{2}\right)^{4/3}\left(\frac{4\pi\rho_0v_{\rm c}H^4}{3t_{\rm cc,diff}^2}\right) \\
    &\approx 6.2 \times10^{37}\,E_{\rm c,50}\rho_{0,-10}^{-1}\theta_{\rm c,10^\circ}^{2/3}H_{14}^{-2}\,{\rm erg}\,{\rm s}^{-1}.
\end{split}
\end{equation}
The cocoon cooling emission would be outshone by the disk emission.

The peak luminosity of a ``kilonova'' is thought to be a factor of $\sim10^3$ higher than a typical nova \citep[e.g.,][]{metzger2010,zhu2020a}. It is expected that the large amount of AGN disk material swept by the ejecta can make the kilonova darker. According to \cite{arnett1982}, the characteristic timescale when the light curve peaks is approximately $t_{\rm kn,p}\approx(\kappa M_{\rm sw,e}/v_{\rm e}c)^{1/2}\approx20\,E_{\rm e,51}^{-1/4}\rho_{0,-10}^{3/4}H_{14}^{9/4}\,{\rm d}$, while the peak luminosity is equal to the heating rate at the peak time i.e., $L_{\rm kn,p} \approx \eta_{\rm th}M_{\rm ej}\epsilon(t_{\rm kn,p}) \approx 10^{40}\,E_{\rm e,51}^{\alpha/4}\rho_{0,-10}^{-3\alpha/4}H_{14}^{-9\alpha/4}(M_{\rm ej}/0.03\,M_\odot)\,{\rm erg}\,{\rm s}^{-1}$, where $\eta_{\rm th}\approx0.5$ is the thermalization efficiency \citep{metzger2010}, the heating rate is $\epsilon(t)\approx2\times10^{10}\,{\rm erg}\,{\rm g}^{-1}\,{\rm s}^{-1}(t/1\,{\rm d})^{-\alpha}$, and $\alpha \approx 1.3$. Therefore, kilonova emission in the AGN accretion disks is also much dimmer than the disk emission.

\subsection{Event Rate and Target-of-opportunity Follow-up of GW triggers}

For a cocoon shock breakout with luminosity $L_{\rm c,bo} \sim10^{46}\,{\rm erg}\,{\rm s}^{-1}$, the detectable distance by EP (field of view $\Omega \approx 1\,{\rm sr}$ and survey sensitivity threshold $F_{\rm th} \approx 10^{-10}\,{\rm erg}\,{\rm s}^{-1}\,{\rm cm}^{-2}$ for a visit) is $D_{\rm L,max} \approx 3,000 \,{\rm Mpc}$, corresponding to a redshift $z_{\rm max} \approx 0.5$. \cite{mckernan2020} showed that the event rate densities of BNS and NSBH mergers from the AGN channel are $R_{\rm BNS}\sim f_{\rm AGN}[0.2,400]\,{\rm Gpc}^{-3}\,{\rm yr}^{-1}$ and $R_{\rm NSBH}\sim f_{\rm AGN}[10, 300]\,{\rm Gpc}^{-3}\,{\rm yr}^{-1}$, respectively. They also suggested that only $\sim10\%-20\%$ of NSBH mergers in AGN disks can result in tidal disruptions. Since the cocoon shock is trans-relativistic (i.e., $v_{\rm c}\sim0.19c$) when it breaks out, the breakout emission would be nearly isotropic and occupy nearly the $4\pi$ solid angle regardless of the opening angle of the breakout flow. On the other hand, the breakout signal may be suppressed if the jet axis has a very small inclination angle with respect to the AGN disk. By assuming that all BNS and $20\%$ NSBH mergers in AGN disks can produce sGRBs and kilonovae, we expect a detection rate of
\begin{equation}
    R_{\rm bo} \approx f_\theta\frac{\Omega}{4\pi} (R_{\rm BNS} + 0.2R_{\rm NSBH})V_{\rm max} \approx f_\theta f_{\rm AGN}[20 , 4200]\,{\rm yr}^{-1}
\end{equation}
by EP for cocoon and ejecta shock breakouts of BNS and NSBH mergers in AGN disks, where $V_{\rm max} = 4\pi D_{L \rm,max}^3/3$, and the factor $f_\theta < 1$ accounts for the fraction of mergers that make a detectable X-ray shock breakout signal due to the distribution of the inclination angle between the binary merger orbital plane and the disk plane.

GW signals from the BNS and NSBH mergers that are embedded in AGN disks are expected to be detected by GW detectors. \cite{abbott2018prospects} showed that an advanced LIGO detector, after achieving their design sensitivity in O4, would have detection range of a BNS (NSBH) merger at $160-190\,{\rm Mpc}$ ($300-330\,{\rm Mpc}$), while in O5 the BNS (NSBH) detection depth for one GW detector could be up to $330\,{\rm Mpc}$ ($590\,{\rm Mpc}$). The expected detection rate of GW triggers that are associated with AGN disk shock breakout signatures is $R_{\rm bo}\sim f_\theta f_{\rm AGN}[0.3 , 20]\,{\rm yr}^{-1}$ in O4 and $R_{\rm bo}\sim f_\theta f_{\rm AGN}[2, 110]\,{\rm yr}^{-1}$ in O5\footnote{In reality, GW signals are usually detected by the GW detectors network. In such cases, more remote BNS and NSBH mergers can be detected by networking of GW detectors \citep[e.g.,][]{zhu2020b} so that the above detection rates estimation for O4 and O5 are  conservative.}.

The cocoon shock breakout and ejecta shock breakout signals usually occur at $\sim 0. 15\,{\rm d}$ and $\sim0.5\,{\rm d}$ after the GW triggers. The delay times between GW triggers and breakout signals reserve enough preparation times for subsequent EM follow-up observations. Multi-messenger observations of GW signals and the two breakout signals from AGN disks can lend strong support for the formation channel of compact binary coalesences in the AGN disks. 

\section{Conclusions and Discussion} \label{sec:conclusions}

A neutron star (BNS or NSBH) merger in the migration traps of an AGN disk can generate two bright EM signatures, i.e., a cocoon shock breakout transient and an ejecta shock breakout transient. We show that the jet launched from such a system would be choked and its energy is deposited to the cocoon which can be energetic enough to break out. Such systems likely would not produce detectable GRBs. The cocoon shock breakout produces soft X-ray emission which can be easily discovered by EP in the future. The second, ejecta breakout would produce a UV/soft X-ray flash. Due to the contamination of the AGN disk emission, such a signal is most prominent in the NUV band and could be detected by UV transient searches. The two breakout signals peak$\sim 0.15\,{\rm d}$ and $\sim0.5\,{\rm d}$ after the BNS or NSBH merger. The cocoon cooling and kilonova emission signals are typically too faint and would be outshone by the disk emission. Since BNS and NSBH mergers are the search targets for GW detectors, taking advantage of GW triggers for subsequent follow-up observations would be essential to detect these breakout signals, which in turn, will offer support to the formation channel of compact binary mergers in the AGN disks. This is also helpful to test the hypothesis of the BBH population in AGN disks and constrain their fraction.

We have assumed that the jet is launched perpendicular to the disk. In general, the jet orientation in the AGN disks could be random. A cocoon shock driven by an inclined jet would take a longer time to break out from the disk with the emission having a lower temperature and luminosity. We also assumed that the first cocoon shock does not significantly affect the disk environment for the later ejecta shock breakout. This needs to be verified by detailed numerical simulations. If the merger leaves behind a long-lived engine, e.g., a magnetar \citep{yu2013,gao2013} or a BH with continued accretion \citep{ma2018}, the ejecta would receive additional energy up to $\sim O(10^{52})\,{\rm erg}$, which would significantly enhance the ejecta shock breakout signature. If this happens, similar to the cocoon breakout case, the ejecta shock breakout event would also make X-rays\footnote{Other X-ray EM counterparts for BNS mergers have been proposed in the past \citep{zhang2013,sun2017}, which involves a post-merger massive magnetar. The prediction seems to match some X-ray transients \citep[e.g.][]{xue2019,sun2019} well. The signals discussed in this paper provide an alternative source to produce bright X-ray transients with a longer timescale that are associated with neutron star  mergers.}.

In the AGN disks, besides compact binary mergers, other destructive explosions, such as supernovae and tidal disruptions \citep[e.g.,][]{assef2018}, could also occur. Shocks would form if these events are embedded in the AGN accretion disk and release a large amount of energy to accelerate AGN disk materials. These events would produce shock breakout events similar to (but probably more energetic than) the transients discussed in this paper.

After the completion of this work, we learned that \cite{perna2020} have independently investigated the general properties of GRBs in an AGN disk. Different from our approach to focus on the characteristics of EM signals of BNS/NSBH mergers located at migration traps in AGN disks, they discussed the various possible outcomes of a GRB originating from different distances from the supermassive BH and different BH masses. For a GRB occurring near the migration trap of an accretion disk, if $M_{\rm SMBH}\gtrsim10^{7}\,M_\odot$, their results showed that the prompt radiation and afterglow emission would occur inside the disk photosphere and get thermalized by Thomson scattering. The outcome would be a thermal transient emerging on a diffusion timescale of days to years. For the case with a smaller $M_{\rm SMBH}$, they indicated the diffusion for prompt and afterglow emissions could be rapid while the late normal afterglow could be observed. This is in general consistent with our conclusions drawn in this paper.

\acknowledgments

We thank an anonymous referee for constructive suggestions. We thank Rosalba Perna for exchanging their paper with us before submission and a referee for constructive suggestions. We thank Yuan-Pei Yang, Liang-Duan Liu, Zhuo Li, Hui Sun, Shao-Ze Li, Yuan-Qi Liu, Er-Lin Qiao for valuable comments. The work of J.P.Z is partially supported by the National Science Foundation of China under Grant No. 11721303 and the National Basic Research Program of China under grant No. 2014CB845800. Y.W.Y is supported by the National Natural Science Foundation of China under Grant No. 11822302, 1183303. H.G. is supported by the National Natural Science Foundation of China under Grant No. 11722324, 11690024, 11633001, the Strategic Priority Research Program of the Chinese Academy of Sciences, Grant No. XDB23040100 and the Fundamental Research Funds for the Central Universities.

\bibliography{BNS_in_AGN}{}
\bibliographystyle{aasjournal}

\end{document}